\documentclass{ws-ijmpd}

\begin{document}

\markboth{Pican\c{c}o, Malheiro \& Ray} {Charged polytropic stars
and a generalization of ...}

\catchline{}{}{}{}{}
%

\title{Charged Polytropic Stars and a Generalization of Lane-Emden Equation}

\author{R. Pican\c{c}o$^\dagger$, M. Malheiro$^\ddagger$ and S. Ray\footnote{Present Address: Inter
University Centre for Astronomy and Astrophysics, Post Bag 4, Pune
411007, India }}

\address{Instituto de F\'{\i}sica, Universidade Federal Fluminense, Av.
Litor\^{a}nea - Boa Viagem\\
Niter\'{o}i, Rio de Janeiro, Brasil\\
$\dagger$rodrigo@if.uff.br, ~$\ddagger$mane@if.uff.br}

\maketitle

\begin{abstract}
In this paper we will discuss charged stars with polytropic
equation of state, where we will try to derive an equation
analogous to the Lane-Emden equation. We will assume that these
stars are spherically symmetric, and the electric field have only
the radial component. First we will review the field equations for
such stars and then we will proceed with the analog of the
Lane-Emden equation for a polytropic Newtonian fluid and their
relativistic equivalent (Tooper, 1964)\cite{1}. These kind of
equations are very interesting because they transform all the
structure equations of the stars in a group of differential
equations which are much more simple to solve than the source
equations. These equations can be solved numerically for some
boundary conditions and for some initial parameters. For this we
will assume that the pressure caused by the electric field obeys a
polytropic equation of state too.
\end{abstract}
\vskip0.1cm
\section*{}
It is known that the structure equations of a Newtonian star (the
hydrostatic equilibrium  and the mass continuity equation) can be
simplified to a second order differential equation of
dimensionless parameters depending on the polytropic index. This
equation is much more simple and elegant than the structure
equations. Once we get to a solution of the differential
equation, then changing just the variables, we can get other
solutions and structure data of our particular interest. This
procedure was first introduced by Lane-Emden and then further
developed by Chandrasekhar. In 1968, Tooper\cite{1} deduced a
relativistic analogous to the Lane-Emden equation with all
corrections impose by General Relativistic. Polytropic charged
stars have been recently discussed in Ray et al.\cite{2} where the
charge distribution was assumed to vary with the mass
distribution. It was showed\cite{2} that in these stars the
electric field must be huge (~$10^{21}$V/m), causing a extremal
instability in the star, and help the star collapse further to a
charged black hole. It will be interesting to get an analogous of
the Lane-Emden equation from such stars. We will try to follow the
Lane-Emden and Tooper steps to get to this equation.
\section{Field Equations}

The metric used here is the usual Schwarzschild metric,
\begin{equation}
ds^2=e^{\nu(r)}c^2dt^2-e^{\lambda(r)} dr^2-r^2(d\theta^2+sen^2\theta
d\phi^2)
\end{equation}
The time-independent gravitational field equations reduces to:
\begin{eqnarray}
e^{-\lambda}(-\frac{1}{r^{2}}+\frac{1}{r}\frac{d\lambda}{dr})+\frac{1}{r^{2}}=-\frac{8\pi
\label{fe1}
G}{c^{4}}(\rho c^{2}+ \frac{E^{2}}{8\pi}) \\
e^{-\lambda}(\frac{1}{r}\frac{d\nu}{dr}+\frac{1}{r^{2}})-\frac{1}{r^{2}}=\frac{8\pi
G}{c^{4}}(p-\frac{E^{2}}{8\pi}) \label{fe2}
\end{eqnarray}
The mixed energy-momentum tensor will include the terms of the
Electromagnetic Field, giving us an equation of the form
\begin{equation}
T_{\nu}^{\mu}=diag(-\rho
c^2-\frac{E^2}{8\pi},p-\frac{E^2}{8\pi},p+\frac{E^2}{8\pi},p
+\frac{E^2}{8\pi}), \label{emt}
\end{equation}
where $p$ is the pressure, $\rho$ the mass density and $E$ the
radial electric field. The four-divergence of $T_{\nu}^{\mu}$
should vanish for it being a conserved quantity and so we get
\begin{equation}
\frac{dp}{dr} = -\frac{1}{2}\frac{d\nu}{dr}(p+\rho
c^2)+\frac{E}{8\pi}(\frac{dE}{dr}+\frac{2E}{r}).\label{eq:dp}
\end{equation}
This is the Tolman-Oppeheimer-Volkoff  (TOV) equation for a
spherically symmetric charged fluid. This equation represents the
hydrostatic equilibrium for the fluid. The first term of the right
hand side (r.h.s.) of equation (\ref{eq:dp}) is derived from the
normal matter-energy, but the second term is new. It represents
the contribution from the coulombian force and the energy from the
electric field. If we look at the energy-momentum tensor
(equation (\ref{emt})), we see that the effective pressure and
density in the fluid is given for
\begin{equation}
p_{ef}=p-\frac{E^2}{8\pi}, \quad \rho_{ef}=\rho+\frac{E^2}{8\pi}
\end{equation}

Now the energy-momentum tensor takes the following form:
\begin{equation}
T_{\nu}^{\mu}=diag(-\rho_{ef} c^2,p_{ef},p+\frac{E^2}{8\pi},p
+\frac{E^2}{8\pi})
\end{equation}

We could have written the last two terms as function of the
effective pressure and density, but we didn't because it is not
advantageous . Now we have the modified TOV equation:
\begin{equation}
\frac {dp_{ef}} {dr}= -\frac{1}{2}\frac{d\nu}{dr}(p_{ef}+\rho_{ef}
c^2)+\frac{E^2}{2\pi r}  \label{TOVm}
\end{equation}

\section{The Generalization of Lane-Emden Equation}

We have from the exterior Schwarzschild solution that
\begin{equation}
e^{-\lambda}=1-\frac{2GM}{rc^2}+\frac{GQ^2}{r^2c^4}. \label{enu}
\end{equation}
Usually we would have only the first term (of the mass) but here
the total mass that an observer really see must include the
contribution of the energy from the electric field. This one is
not limited to the surface of the stars and is observed by a
distant observer, so we have to include the contribution of the
electric energy seen from infinity. So we can write
\begin{equation}
M=\int_{0}^{+\infty} 4\pi r^2(\rho + \frac{E^2}{8\pi c^2})dr
,\quad Q=\int_{0}^{R}4\pi r^2 \rho_{ch}e^{\lambda /2} dr,
\end{equation}
where $\rho_{ch}$ is the charge density. Now we can define an
effective mass given by
\begin{equation}
M_{ef}=M-\frac{Q^2}{2\pi rc^2}. \label{mef1}
\end{equation}
With the help of the equations (\ref{enu}) and (\ref{mef1}) we
define $x$ as
\begin{equation}
x=\frac{(1-e^{-\lambda})rc^2}{2GM_{ef}}.
\end{equation}
With this we rewrite $e^{-\lambda}$ in the following form:
\begin{equation}
e^{-\lambda}=1-\frac{2GM_{ef}x}{rc^2}. \label{ex}
\end{equation}
Writing equation (\ref{fe1}) as function of $x$ and using equation
(\ref{ex}) we get the following relation:
\begin{equation}
M_{ef}\frac{dx}{dr}=4\pi r^2 \rho_{ef}. \label{mx}
\end{equation}
Now we have to write the explicit relation between effective
pressure and effective density, then use in the modified TOV
equation (\ref{TOVm}). We will assume that this relation is given
by the following polytropic equation:
\begin{equation}
p_{ef}=K_{efe}\rho_{ef}^{1+1/n}.
\end{equation}
If we do a dimensional analysis we will see that all terms of the
r.h.s. have the dimension of pressure divided by unit of length,
which is not surprising since the left hand side is a derivative
of the pressure with respect to the radius. In this sense we will
call the second term of the r.h.s. in the following form :
\begin{equation}
\frac{E^2}{2\pi r}=\frac{dp_{el}}{dr}
\end{equation}
where $p_{el}$ is a pressure related to the electric field. And in
order to make further progress we will suppose that this pressure
also obeys a polytropic equation of state
\begin{equation}
p_{el}=K_{el}\rho_{el}^{1+1/n}.
\end{equation}
Now let us write the density in a parametric form
\begin{equation}
\rho_{ef}=\rho_{efc} \theta^{n}, \quad \rho_{el}=\rho_{elc}
\theta^{n},
\end{equation}
where $\rho_{efc}$ is the central effective pressure and
$\rho_{elc}$ is the central electric pressure. In this notation,
both electric and effective pressure take the form
\begin{eqnarray}
p_{ef}=K_{efe}\rho _{efc}^{1+1/n}\theta^{n+1}, \quad
p_{el}=K_{el}\rho _{elc}^{1+1/n}\theta^{n+1}.
\end{eqnarray}
Using these modifications on equation (\ref{TOVm}) we get
\begin{equation}
\frac{d\nu}{dr}=-2(n+1)\frac{(\sigma -\sigma_{el} \eta)}{\sigma
\theta +1}\frac{d\theta}{dr}, \label{tetanu}
\end{equation}
where
\begin{equation}
\sigma = \frac{K\rho_{efc}^{1/n}}{c^2}, \quad \sigma_{el}
=\frac{K\rho_{elc}^{1/n}}{c^2}, \quad \eta =
\frac{\rho_{elc}}{\rho_efc}.
\end{equation}
Integrating equation (\ref{tetanu}) and using the fact that
$\nu\rightarrow \nu_c$ when $\theta \rightarrow 1$ we get
\begin{equation}
e^{\nu}=e^{\nu_c}(\frac{\sigma +1}{\sigma \theta
+1})^{2(\frac{n+1}{\sigma})(\sigma -\sigma_{el} \eta)}.\label{nuf}
\end{equation}

Using equation (\ref{nuf}), when $r=R$ we have $\rho\rightarrow 0$
and $\theta = 0$, and comparing equation (\ref{enu}) we get

\begin{equation}
e^{\nu}=(\sigma \theta +1)^{-2 \frac{n+1}{\sigma}(\sigma
-\sigma_{el} \eta)}(1 -\frac{2GM_{ef}}{rc^2}).
\end{equation}

Using equation (\ref{tetanu}) together with the  field equation
(\ref{fe2}), we get a new differential equation connecting $x$ and
$\theta$
\begin{equation}
\frac{(n+1)(\sigma -\sigma_{el}\eta)}{\sigma \theta
+1}r\frac{d\theta}{dr}(1-\frac{2GM_{ef}x}{rc^2})+\frac{GM_{ef}x}{rc^2}+\frac{G\sigma
M_{ef}}{c^2}\frac{dx}{dr}\theta =0.
\end{equation}

Using this together with equation (\ref{mx}) and the following
changing of variables
\begin{equation}
r=\frac{\xi}{A}, \quad v(\xi)=\frac{A^{3}M_{ef}x}{4\pi
\rho_{ef}}, \quad A=[\frac{4\pi G
\rho_{efc}}{(n+1)K\rho_{efc}^{1/n}}]^{1/2},
\end{equation}
we get the following differential equations:
\begin{equation}
\xi
^{2}\frac{d\theta}{d\xi}\frac{(1-\frac{\sigma_{el}}{\sigma}\eta)-(n+1)(\sigma-\sigma_{el}\eta)v/\xi}{\sigma
\theta +1}+v+\sigma\theta\xi \frac{dv}{d\xi} =0;
\end{equation}
\begin{equation}
\frac{dv}{d\xi}=\xi^2 \theta^n.
\end{equation}

These are the General Lane-Emden equation for a spherically
symmetric charged polytropic fluid. Solving these equation we can
step back and find those variables of our interest. Numerical
solution and verification of these equations with the results
obtained by solving the TOV in the {\it normal} procedure is
beyond the scope of this present paper and will be shown in a
future article.

\end{document}